\documentclass[aip,reprint]{revtex4-1}

\draft 
\usepackage{graphicx} 
\usepackage{times} 
\usepackage{amsmath} 
\usepackage{amssymb}  
\usepackage{bm}
\usepackage{algorithm}
\usepackage{algorithmic}
\begin{document}


\title{Model-free Control of Chaos with Continuous Deep Q-learning} 



\author{Junya Ikemoto}%
\email{ikemoto@hopf.sys.es.osaka-u.ac.jp}
\affiliation{Graduate School of Engineering Science, Osaka University\\
Toyonaka, Osaka, 560-8531, Japan\\}
\author{Toshimitsu Ushio}%
\email{ushio@sys.es.osaka-u.ac.jp}
\affiliation{Graduate School of Engineering Science, Osaka University\\
Toyonaka, Osaka, 560-8531, Japan\\}


\date{\today}

\begin{abstract}
The OGY method is one of control methods for a chaotic system. In the method, we have to calculate a stabilizing periodic orbit embedded in its chaotic attractor. Thus, we cannot use this method in the case where a precise mathematical model of the chaotic system cannot be identified. In this case, the delayed feedback control proposed by Pyragas is useful. However, even in the delayed feedback control, we need the mathematical model to determine a feedback gain that stabilizes the periodic orbit. To overcome this problem, we propose a model-free reinforcement learning algorithm to the design of a controller for the chaotic system. In recent years, model-free reinforcement learning algorithms with deep neural networks have been paid much attention to. Those algorithms make it possible to control complex systems. However, it is known that model-free reinforcement learning algorithms are not efficient because learners must explore their control policies over the entire state space. Moreover, model-free reinforcement learning algorithms with deep neural networks  have the disadvantage in taking much time to learn their control optimal policies. Thus, we propose a data-based control policy consisting of two steps, where we determine a region including the stabilizing periodic orbit first, and make the controller learn an optimal control policy for its stabilization. In the proposed method, the controller efficiently explores its control policy only in the region.
\end{abstract}


\maketitle 
\begin{quotation}
In general, periodic orbits embedded in chaotic attractors depend on the parameters of the chaotic system and the chaos control method that does not need the precise computation of the orbit is practically useful. Several such methods such as delayed feedback control have been proposed.  However, in these methods, the identification of the parameters are required. Thus, we propose a model-free control method using continuous deep Q-learning. Continuous deep Q-learning is one of the deep reinforcement leaning algorithms and has been applied to controls of complex tasks recently. We propose a reward that evaluates stabilization by the control inputs. Moreover, since the stabilized periodic orbit is embedded in a chaotic attractor, we select a region including the orbit where we inject the control inputs so that efficient learning is achieved. As example, we consider stabilization of a fixed point embedded in a chaotic attractor of the Gumowski-Mira map and it is shown by simulation that we learn a nonlinear state feedback controller by the proposed method.
\end{quotation}
\section{Introduction}
It is known that many unstable periodic orbits are embedded in chaotic attractors. Using this property, Ott, Grebogi, and Yorke proposed an efficient chaos control method \cite{OGY}. However, when we use this method, we have to calculate a stabilizing periodic orbit embedded in the chaotic attractor precisely. In the case where we cannot identify precise mathematical models of the chaotic systems, the delayed feedback control \cite{Pyragas} is known to be very useful. Many related methods have been proposed \cite{Ushio_1,Nakajima_1,Extend_delay,Yamamoto,Nakajima_2}. Moreover, the  prediction-based chaos control method using predicted future states was also proposed \cite{Ushio_2}. However, it is difficult to determine a feedback gain of the controller in the absence of its mathematical model. To overcome this problem, a method of adjusting the gain parameter using the gradient method was proposed \cite{Nakajima_3}. Neural networks have been used as model identification \cite{Boukabou,Shen}. Reinforcement Learning (RL) has been also applied to the design of the controller \cite{Der,Der2,Funke,Der3,Randlov,Gadaleta1,Gadaleta2}. Recently, RL with deep neural networks, which is called Deep Reinforcement Learning (DRL), has been paid much attention to. DRL makes it possible to learn better policies than human level policies in Atari video games \cite{DQN} and Go \cite{Silver}. DRL algorithms have been applied not only to playing games but also to controlling real-world systems such as autonomous vehicles and robot manipulators. As an application of the physics field, the control method of a Kuramoto-Sivashinsky equation, which is one-dimensional time-space chaos, using the DDPG algorithm \cite{DDPG} was proposed \cite{Bucci}. 

In this paper, we apply a DRL algorithm to the control of chaotic systems without identifying their mathematical model. However, in model-free RL algorithms, the learner has to explore its optimal control policy over the entire state space, which leads to inefficient learning. Moreover, when we use deep neural networks, it takes much time for the learner to optimize many parameters in the deep neural network. In this paper, we propose an efficient model-free control method consisting of two steps. First, we determine a region including a stabilizing periodic orbit based on uncontrolled behavior of the chaotic systems.  Next, we explore an optimal control policy in the region using deep Q networks while we do not control the system outside the region.  Without loss of the generality, we focus on the stabilization of a fixed point embedded in the chaotic attractor.

This paper organizes as follows. In Section I\hspace{-1pt}I, we show a method to determine a region including a stabilizing fixed point. In Section I\hspace{-1pt}I\hspace{-1pt}I, we propose a model-free reinforcement learning method to explore an optimal control policy in the region. In Section I\hspace{-1pt}V, numerical simulations of the proposed chaos control of the Gmoowski-Mira map, which is an example of discrete-time chaotic system, is performed to show the usefulness of the proposed method. Finally, in Section V, we describe the conclusion of this paper and future work.

\section{Estimation of region}
We consider the following chaotic discrete-time system.
\begin{eqnarray}
\bm{x}_{k+1}=F(\bm{x}_{k},\bm{u}_{k}),
\end{eqnarray}
where $\bm{x}\in\mathbb{R}^{n}$ is the state of the chaotic system and $\bm{u}\in\mathbb{R}^{m}$ is the control input. We assume that the function $F$ cannot be identified precisely. Thus, we cannot calculate a precise value of the stabilizing periodic orbits embedded in its chaotic attractor. On the other hand, although the state of the chaotic system does not converge to the periodic orbit, it is sometimes close to any unstable periodic orbit embedded in the chaotic attractor. Using this property, we observe the behavior of the chaotic system without the control input and sample states that are close to the stabilizing periodic orbit. In the following, for simplicity, we focus on the stabilization of a fixed point embedded in the chaotic attractor.  We observe behaviors $\bm{x}_k\ (k=0,1,\ldots)$ of the uncontrolled chaotic system ($\bm{u}_k=0$) and sample states $\bar{\bm{x}}^{(l)}=\bm{x}_{k_{l}}\ (l=1,2,...,L)$ satisfying the following condition from the behaviors, where $L$ is the number of the sampled states.
\begin{eqnarray}
\| \bm{x}_{k_{l}+1} - \bm{x}_{k_{l}}\|_{p} < \epsilon,
\end{eqnarray}
where $\| \cdot \|$ denotes the $\ell_p$-norm over $R^n$ and $\epsilon$ is a sufficiently small positive constant. We estimate the stabilizing fixed point $\hat{\bm{x}}_{f}$ based on $L$ sampled states $\bar{\bm{x}}^{(l)}$.
\begin{eqnarray}
\hat{\bm{x}}_{f}=\frac{1}{L}\sum^{L}_{l=1}\bar{\bm{x}}^{(l)}\ \ \ \ \ (l=1,2,...,L).
\end{eqnarray}
Note that there may exist more than one fixed point in the chaotic attractor in general.  In such a case, we calculate clusters of the sampled data corresponding to the fixed points and select the cluster close to the stabilizing fixed point.

\begin{figure}
  \centering
　 \includegraphics[width=7.5cm]{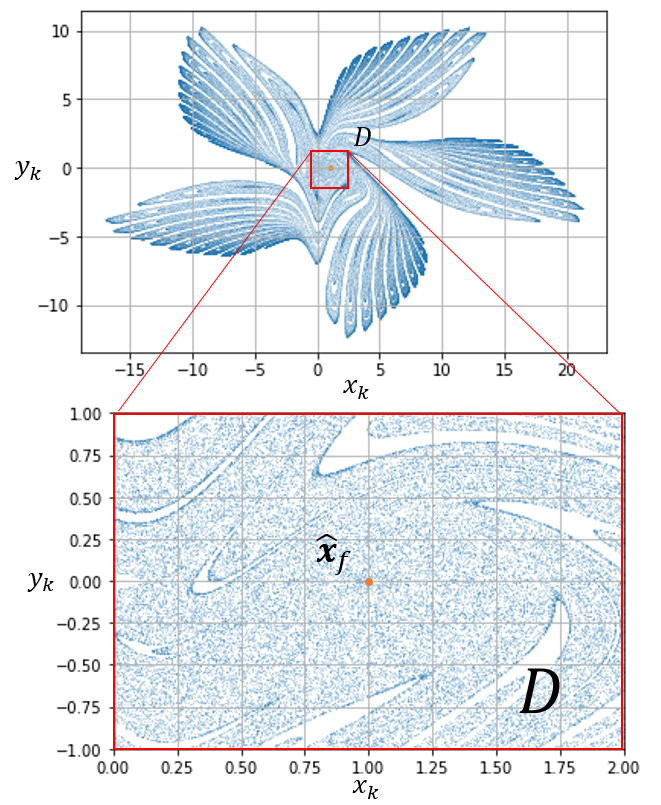}
	\caption{Example of the region $D$ for the Gumowski-Mira map. The estimated fixed point is $\hat{\bm{x}}_{f}=[0.994,0.001]^{T}$. We set the region $D=\{(x,y)|\ ||\bm{x}-\hat{\bm{x}}_{f}||_{\infty}\le1\}$.\label{Example}}
\end{figure}

Then, we set a region $D$ appropriately based on the estimated fixed point $\hat{\bm{x}}_{f}$, where the center of $D$ is the estimated fixed point $\hat{\bm{x}}_{f}$. We have to select the region enough large that the stabilizing fixed point is sufficiently far from the boundary of the region. Since we use a deep neural network, we can make a learner learn a nonlinear control policy for a large region while both the OGY method and the delayed feedback control method are linear control methods. As an example, we show an estimation of the fixed point of the Gumowski-Mira map in Fig.\ \ref{Example}.

Furthermore, we transform the state $x$ into the following new state $\bm{s}\in \mathcal{S}$. 
\begin{eqnarray}
\bm{s}:=\phi(\bm{x}),
\end{eqnarray}
where $\phi:\mathbb{R}^{n}\to \mathcal{S}$ is the following coordinate transformation. 
\begin{eqnarray}
\phi(\bm{x}):=\begin{cases}
    \bm{x}-\hat{\bm{x}}_{f}\ \ \ \ \bm{x} \in D \\
    \bm{s}_{out}\ \ \ \ \bm{x}\notin D
  \end{cases}.
\end{eqnarray}
The transformed state space $\mathcal{S}$ is $D'\cup\{\bm{s}_{out}\}$, where $D'=\{\phi(\bm{x})|\bm{x}\in D \}$. The state $\bm{s}_{out}$ represents that the current state of the chaotic systems lies out of the region $D$ so that the control input is set to 0 and we do not sample the state for learning. Then, the origin of the state space $D'$ coincides with the estimated fixed point $\hat{\bm{x}}_{f}$.

\section{Deep Reinforcement Learning for Chaos Control}
The goal of RL is to learn an optimal control policy in the long run through interactions between a controller with a learner and a system. First, the controller observes the system state $\bm{x}$ and computes the control input $\bm{u}$ in accordance with its control policy $\mu$．Next, the controller inputs the control input $\bm{u}$ to the system and the state of the system moves from $\bm{x}$ to $\bm{x}'$. Finally, the controller observes the next state $\bm{x}'$ and receives the immediate reward $r$. The immediate reward is determined by the reward function $R$.
\begin{figure}
  \centering
　 \includegraphics[width=8.5cm]{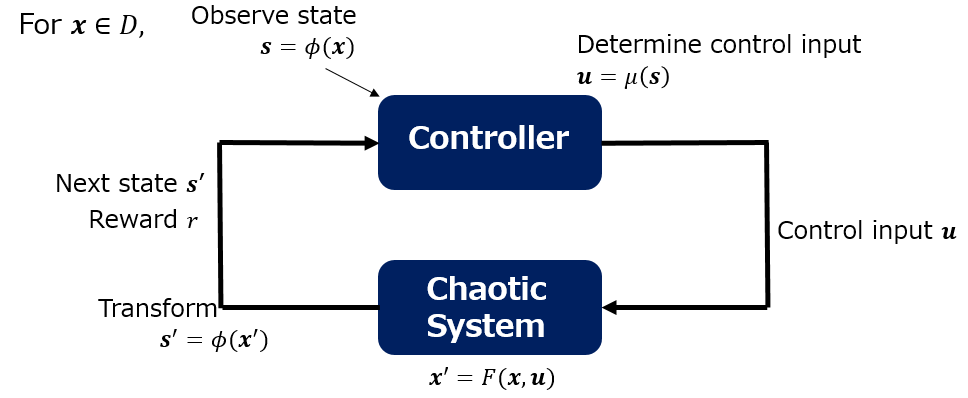}
	\caption{Interactions between a system and a controller. In this paper, we regard the transformed state $\bm{s}\in S$ as the state in the RL framework. The controller observes the transformed state $\bm{s}$ and computes the control input $\bm{u}$ in accordance with its policy $\mu$. The controller inputs the controller input $\bm{u}$ to the system and the state of the system moves from $\bm{s}$ to $\bm{s}'$. Finally, the controller observes the next transformed state $\bm{s}'$ and receives the immediate reward $r$. The controller updates its control policy $\mu$ based on the transition $(\bm{s},\bm{u},\bm{s}',r)$.\label{RL}}
\end{figure}
In this paper, we make the controller learn its control policy only in the region $D$ to improve its learning efficiency. Thus, we define $\bm{s}\in\mathcal{S}$ as the state of the RL framework. Interactions between them is shown in Fig.\ \ref{RL}. 

In this paper, the reward function $R:D'\times\mathbb{R}^{m}\times\mathcal{S}\to\mathbb{R}$ is defined by
\begin{eqnarray}
R&&(\bm{s},\bm{u},\bm{s}')=\nonumber\\
&&\begin{cases}
    -(\bm{s}'-\bm{s})^{T}M_{1}(\bm{s}'-\bm{s})-\bm{u}^{T}M_{2}\bm{u} \ \ \ \ \ \mbox{if}\  \bm{s}' \neq \bm{s}_{out}\\
    -q\ \ \ \ \ \mbox{otherwise},
  \end{cases}\nonumber\\
\end{eqnarray}
where $M_{1}$ and $M_{2}$ are positive definite matrices and $q$ is a sufficiently large positive constant. Since $\hat{\bm{x}}_{f}$ is an approximation of the fixed point, the controller requires exploring the fixed point through its learning. Thus, we define the reward function $R$ that takes the maximum reward when the state of the system is stabilized at the fixed point $\bm{x}_{f}$. In the case of $\bm{s}'=\bm{s}_{out}$, the reward function takes the sufficiently large penalty $-q$. Moreover, since the goal of RL is to learn the control policy that maximizes the long-term reward, we define the following value functions.
\begin{eqnarray}
V^{\mu}(\bm{s})&=&\mathbb{E}\left[ \sum^{\infty}_{n=i} \gamma^{n-i}r_{n}|\bm{s}_{i}=\bm{s}\right], \\
Q^{\mu}(\bm{s},\bm{u})&=&\mathbb{E}\left[ \sum^{\infty}_{n=i} \gamma^{n-i}r_{n}|\bm{s}_{i}=\bm{s},\bm{u}_{i}=\bm{u}\right], 
\end{eqnarray}
where $\gamma\in[0,1)$ is a discount rate to prevent divergences of the value functions. Eqs. (7) and (8) are called a state value function and a state-action value function (Q-function), respectively. These value functions represent the mean of the discounted sum of immediate rewards which the controller receives in accordance with its control policy $\mu$, where we do not include immediate rewards in Eqs. (7) and (8) after the state of the system moves to $\bm{s}_{out}$, that is, the transformed state $\bm{s}_{out}$ is a termination state for a learning episode.

\begin{figure*}
  \centering
　 \includegraphics[width=16.0cm]{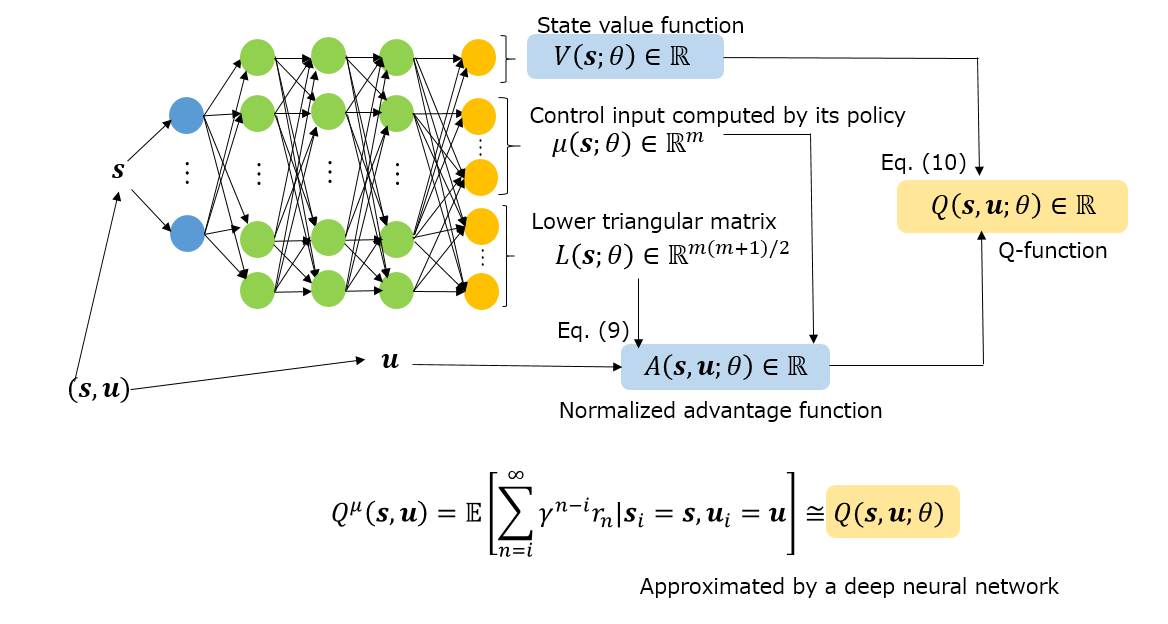}
	\caption{Illustration of the deep neural network for the continuous deep Q-learning algorithm. The input to the deep neural network is the transformed state $\bm{s}$ and outputs are the approximated state value function $V(\bm{s};\theta)$, the control input $\mu(\bm{s};\theta)$, and elements of the lower triangular matrix $P_{L}(\bm{s};\theta)$. We define the normalized advantage function (NAF) as Eq.\ (9). Moreover, by adding the NAF and the approximated state value function, we approximate the Q-function. Note that $Q(\bm{s},\bm{u};\theta)=V(\bm{s};\theta)$ when the approximated Q-function $Q(\bm{s},\bm{u};\theta)$ is maximized for the control input $\bm{u}$.\label{DNN}}
\end{figure*}

Furthermore, we apply DRL to design the controller. In DRL, the control policy function and value functions are approximated by deep neural networks. DDPG \cite{DDPG} and A3C \cite{A3C} are DRL algorithms for continuous control problems. However, it is difficult to handle these algorithms because the control policy function and value functions are approximated by separate deep neural networks in these algorithms. On the other hand, in a continuous deep Q-learning algorithm \cite{NAF}, we can approximate the control policy function and value functions by only one deep neural network. Thus, in this paper, we use the continuous deep Q-learning algorithm. The illustration of the deep neural network used in the algorithm is shown in Fig.\ \ref{DNN}, where $\theta$ is the parameter vector of the deep neural network.
\begin{figure*}
  \centering
　 \includegraphics[width=16.0cm]{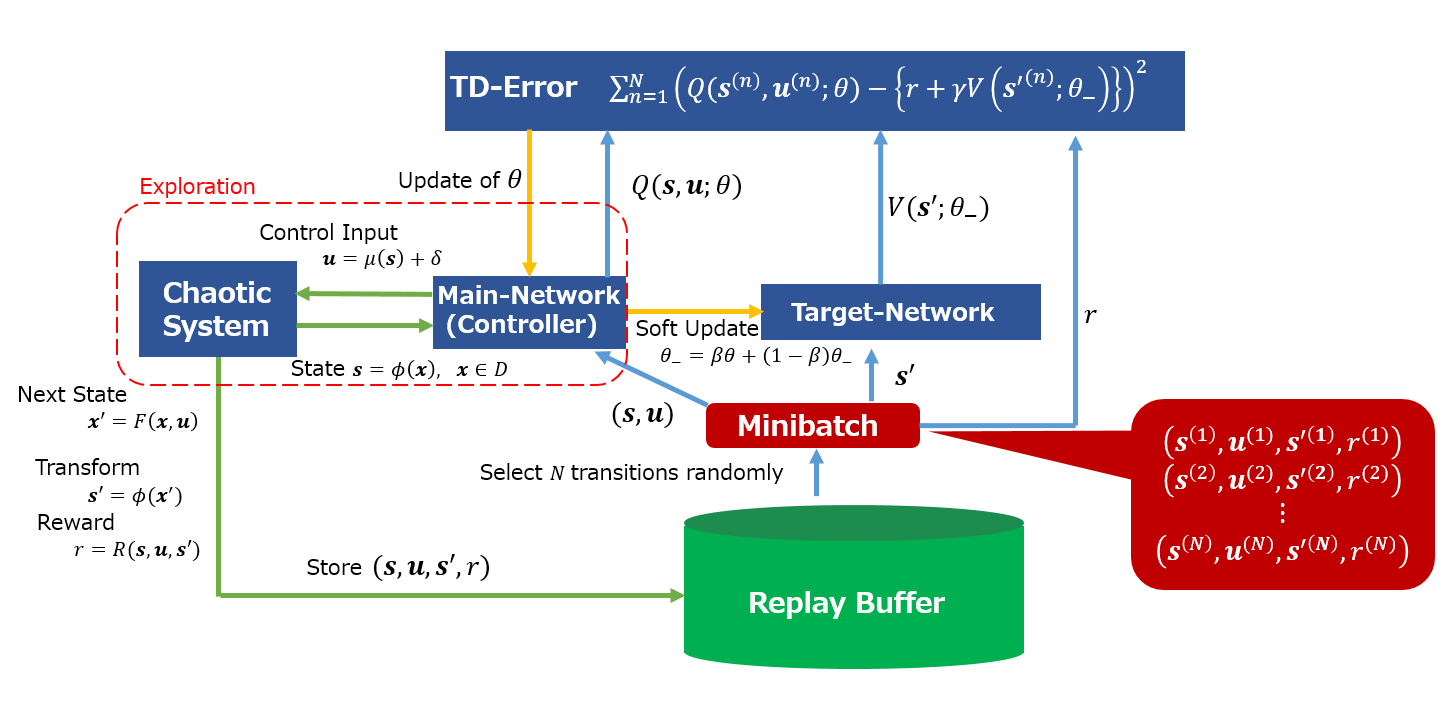}
	\caption{Illustration of controlled chaotic systems by the proposed learning controller. The chaotic system and the main-network keep generating transitions $(\bm{s},\bm{u},\bm{s}',r)$, where $\bm{s}$, $\bm{u}$, $\bm{s}'$, and $r$ are the transformed state of the chaotic system, the control input, the next transformed state of the chaotic system, and the immediate reward. The transition $(\bm{s},\bm{u},\bm{s}',r)$ is stored in the replay buffer $B$. At the time of updating the parameter vector of the deep neural network $\theta$, $N$ transitions $(\bm{s}^{(n)},\bm{u}^{(n)},\bm{s}'^{(n)},r^{(n)})\ (n=1,2,...,N)$ are randomly selected to make a minibatch. The parameter vector $\theta$ is updated based on the minibatch. On the other hand, the parameter vector of the target network $\theta_{-}$ is updated by $\theta_{-} \gets \beta\theta + (1-\beta)\theta_{-}$.\label{Algorithm}}
\end{figure*}
The input to the deep neural network is the transformed state $\bm{s}$ and outputs are the approximated state value function $V(\bm{s};\theta)$, the control input $\mu(\bm{s};\theta)$, and elements of the lower triangular matrix $P_{L}(\bm{s};\theta)$ with the diagonal terms exponentiated. We define the normalized advantage function (NAF) as follows.
\begin{eqnarray}
A(\bm{s},\bm{u};\theta)=\hspace{6cm} \nonumber\\
-\frac{1}{2}(\bm{u}-\mu(\bm{s};\theta))^{T}P_{L}(\bm{s};\theta)P_{L}(\bm{s};\theta)^{T}(\bm{u}-\mu(\bm{s};\theta)),\nonumber\\
\end{eqnarray}
where $\bm{u}$ is the control input to the system at the transformed state $\bm{s}$. Note that $P_{L}(\bm{s};\theta)P_{L}(\bm{s};\theta)^{T}$ is the positive definite matrix because $P_{L}(\bm{s};\theta)$ is the lower triangular matrix. Therefore, the maximum value of the NAF with respect to the control input $\bm{u}$ is 0. Then, the control input $\bm{u}=\mu(\bm{s};\theta)$. Eq. (9) is the quadratic approximation of the advantage function \cite{Sutton} that represents how much the control input $\bm{u}$ is superior to the control input computed in accordance with the policy $\mu$. 

By adding the NAF and the approximated state value function, we approximate the Q-function as follows.
\begin{eqnarray}
Q(\bm{s},\bm{u};\theta)&=&V(\bm{s};\theta)+A(\bm{s},\bm{u};\theta).
\end{eqnarray}
We describe the learning method. We define the following TD-error to update the parameter vector of the deep neural network. 
\begin{eqnarray}
J(\theta)&=&\mathbb{E}\left[ (Q(\bm{s},\bm{u};\theta)-(r+\gamma\max_{\bm{u}'}Q(\bm{s}',\bm{u}';\theta)))^{2}\right]\nonumber\\
&=&\mathbb{E}\left[ (Q(\bm{s},\bm{u};\theta)-(r+\gamma V(\bm{s}';\theta)))^{2}\right],
\end{eqnarray}
where $V(\bm{s}_{out};\theta)=0$. The parameter vector $\theta$ is updated to the direction of minimizing the TD-error using an optimizing algorithm such as Adam \cite{Adam}.

In the learning, we use a target network \cite{DQN}, which is another deep neural network, to update the parameter vector $\theta$, where the parameter vector of the target network is denoted by $\theta_{-}$.  When we compute the approximated state value function $V(\bm{s}';\theta)$ in Eq.\ (11), we use the output of the target network as follows. 
\begin{eqnarray}
J(\theta)=\mathbb{E}\left[ (Q(\bm{s},\bm{u};\theta)-(r+\gamma V(\bm{s}';\theta_{-})))^{2}\right].
\end{eqnarray}
The target network prevents the learning from being unstable. The parameter vector $\theta_{-}$ is updated by the following equation. 
\begin{eqnarray}
\theta_{-}=\beta\theta+(1-\beta)\theta_{-},
\end{eqnarray}
where $\beta$ is the learning rate of the target network and set to a sufficiently small positive constant. This update method is called a soft update.

Moreover, we use the experience replay \cite{DQN}. In the experience replay, the controller does not immediately use the transition $(\bm{s},\bm{u},\bm{s}',r)$ obtained by the exploration for its learning. The controller stores the transition in the replay buffer $B$ once and randomly selects $N$ transitions to make a minibatch at the time of the update of $\theta$. The experience replay is a method to remove the correlation of transitions. Note that, since we learn an optimal policy only in the region $D'$, we do not store all behaviors but the transitions in the region.

In the exploration for the optimal control policy, the controller determines the control input as follows. 
\begin{eqnarray}
\bm{u}=\mu(\bm{s};\theta)+\delta,
\end{eqnarray}
where $\delta$ is an exploration noise according with an exploration noise process $\mathcal{N}$ that we properly have to set. 

The whole learning algorithm is shown in Algorithm\ 1 and the controlled chaotic system is illustrated in Fig.\ \ref{Algorithm}. $M$ is the number of behaviors. $K$ is the maximum discrete-time step of one behavior. $I$ is the frequency of the update of $\theta$ per $k_{p}$ discrete-time steps.

\begin{algorithm}[H]               
\caption{Continuous Deep Q-learning for Chaos Control}    
\label{alg2}                          
\begin{algorithmic}[1]
\STATE Initialize the replay buffer $B$.
\STATE Randomly initialize the main Q network with weights $\theta$.
\STATE Initialize the target network with weights $\theta_{-} = \theta$.
\STATE Estimate the fixed point $\hat{\bm{x}}_{f}$ and select $D$.
\FOR{behavior$=1,...,M$}
\STATE Initialize the initial state $\bm{x}_{0}$.
\STATE Initialize a random process $\mathcal{N}$ for action exploration ($\delta\sim\mathcal{N}$).
\FOR{$k=0,...,K$}
\IF{$k\ \%\ k_{p}=0$}
\FOR{iteration$=1,...,I$}
\STATE Sample a random minibatch of $N$ transitions $(\bm{s}^{(n)},\bm{u}^{(n)},\bm{s}'^{(n)},r^{(n)}),\ n=1,...,N$ from $B$.
\STATE Set $t^{(n)}$
\[
        t^{(n)} = \begin{cases}
                  r^{(n)}+\gamma V(\bm{s}'^{(n)};\theta_{-})\ \ \ \ \ \bm{s}'^{(n)} \neq \bm{s}_{out}\\
   			r^{(n)}\ \ \ \ \ \mbox{otherwise}
                \end{cases}
\]
\STATE Update $\theta$ by minimizing the TD error: $J(\theta)=\frac{1}{N}\sum_{n=1}^{N} (Q(\bm{s}^{(n)},\bm{u}^{(n)};\theta)-t^{(n)})^{2}$.
\STATE Update the target network: $\theta_{-} \gets \beta\theta + (1-\beta)\theta_{-}$.
\ENDFOR
\ENDIF
\IF{$\bm{x}_{k} \in D$}
\STATE Transform the observed state $\bm{x}_{k}$ into $\bm{s}=\phi(\bm{x}_{k})$.
\STATE Determine the exploratory action $\bm{u} = \mu(\bm{s};\theta) + \delta$.
\STATE Input $\bm{u}$ to the chaotic system and the state moves to the next state $\bm{x}_{k+1}$.
\STATE Observe the next state $\bm{x}_{k+1}$.
\STATE Transform the observed state $\bm{x}_{k}$ into $\bm{s}'=\phi(\bm{x}_{k+1})$.
\STATE Return the immediate reward $r=R(\bm{s},\bm{u},\bm{s}')$.
\STATE Store the transition $(\bm{s},\bm{u},\bm{s}',r)$ in $B$.
\ELSE 
\STATE The state is transited to the next state $\bm{x}_{k+1}$ without the control input.
\ENDIF
\STATE $\bm{x}_{k+1}\gets\bm{x}_{k}$.
\ENDFOR
\ENDFOR
\end{algorithmic}
\end{algorithm}

\section{Example}
In order to show the usefulness of the proposed method, we perform the numerical simulation of the chaos control of the Gumowski-Mira map \cite{Gumowski-Mira}, which is an example of the discrete-time chaotic system. The Gumowski-Mira map is described by
\begin{eqnarray}
x_{k+1}&=&y_k+b(1-0.05y_{k}^{2})y_{k}+f_1(x_{k})+0.1u_{k},\\
y_{k+1}&=&-x_{k} + f_1(x_{k+1}),
\end{eqnarray}
where $f_1$ is given by
\begin{eqnarray}
f_{1}(x)=\eta x +\frac{2(1-\eta)x^{2}}{1+x^{2}}.
\end{eqnarray}
In this paper, we assume that $b=0.008$ and $\eta=-0.8$, where we cannot use these parameters to design the controller.  

\begin{figure}
  \centering
　 \includegraphics[width=8.5cm]{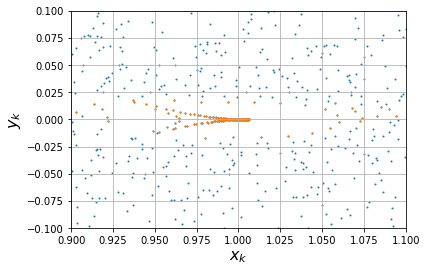}
	\caption{Points of states of the chaotic system without the control input. The orange plots are states which satisfy Eq.\ (2) with $\epsilon=0.02$. We regard the mean of orange points as an estimated fixed point.\label{estimation}}
\end{figure}

By simulations, we observe the uncontrolled behaviors of the chaotic system to estimate the fixed point. We set $\epsilon=0.02$ and $p=1$ ($\ell_1$-norm) in Eq.\ (2). Then, the estimated fixed point is $\hat{\bm{x}}_{f}=[0.994,0.001]^{T}$. Thus, we select the following region $D$ shown in FIG.\ \ref{estimation}.
\begin{eqnarray}
D:=\{(x,y)|-0.006\le x \le1.994,\ -0.999 \le y \le 1.001\}.\nonumber\\
\end{eqnarray}
Then, if $[x,y]^{T}\in D$, the transformed state is $\bm{s}=[s^{x},s^{y}]^{T}=[x-0.994,y-0.001]^{T}$. Otherwise, the transformed state is $\bm{s}=\bm{s}_{out}$.

We use a deep neural network with three hidden layers, where all hidden layers have 32 units and all layers are fully connected layers. The activation functions are ReLU except for the output layer. Regarding the activation functions of the output layer, we use a linear function at both units for the approximated state value function $V(\bm{s};\theta)$ and elements of the matrix $L_{p}(\bm{s};\theta)$, while we use a 2 times weighted hyperbolic tangent function at the units for the control inputs $\mu(\bm{s};\theta)$. The size of the replay buffer is $1.0\times10^{6}$ and the minibatch size is 64. The parameter vector of the deep neural network is updated by ADAM \cite{Adam}, where its stepsize is set to $1.25\times10^{-3}$. The soft update late $\beta$ for the target network is 0.01, and the discount rate $\gamma$ for the Q-function is 0.99. For the exploration noise process $\mathcal{N}$, we use an Ornstein Uhlenbeck process \cite{OUnoise}.

Moreover, we set parameters of the reward function (6) as follows.
\begin{eqnarray}
M_{1}&=&\left[
    \begin{array}{ccc}
      0.08 & 0 \\
      0 & 0.08 
    \end{array}
  \right],\\
M_{2}&=&0.18,\\
q&=&20.0.
\end{eqnarray}
In the simulation, we assume that state transitions of the system occur 10800 times per one behavior ($K=10800$). Moreover, we assume that the parameter vector of the deep neural network $\theta$ is updated twice ($I=2$) every 80 state transitions ($k_{p}=80$). 

\begin{figure}
  \centering
　 \includegraphics[width=9.0cm]{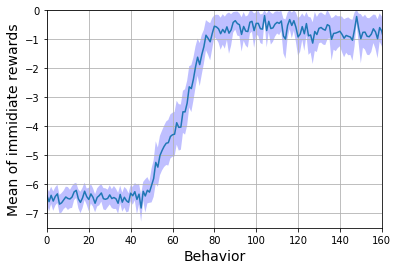}
	\caption{Learning curve. It is the mean of the immediate rewards obtained within 10800 transitions after the each learning episode. The solid line represents the average learning performance obtained in 100 times of learning and the shade represents the 99$\%$ confidence interval.\label{Learning Curve}}
\end{figure}

We show simulation results. The learning curve is shown in Fig.\ \ref{Learning Curve}. The horizontal axis represents the number of episodes and the vertical axis represents the mean value of the immediate rewards obtained within 10800 transitions ($0\le k\le10800$). The solid line represents the average learning performance obtained in 100 times of learning and the shade represents the 99$\%$ confidence interval. It is shown that high immediate rewards are obtained as updates of the parameter vector of the deep neural network are repeated.

\begin{figure}
  \centering
　 \includegraphics[width=9.0cm]{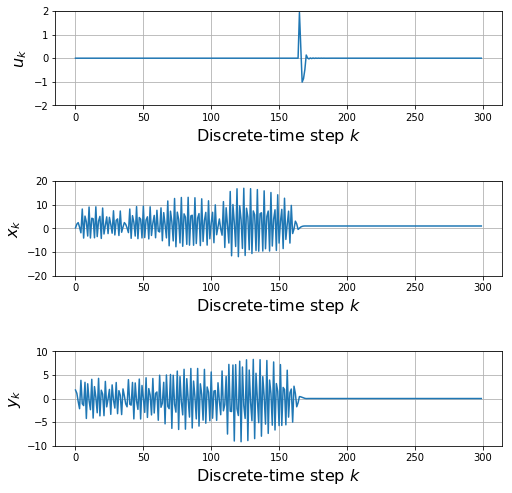}
	\caption{Time response of the chaotic system by the controller that sufficiently learned its control policy, where the initial state $\bm{x}_{0}$ is $[0.2,1.8]^{T}$.\label{response} }
\end{figure}

\begin{figure}
  \centering
　 \includegraphics[width=8.5cm]{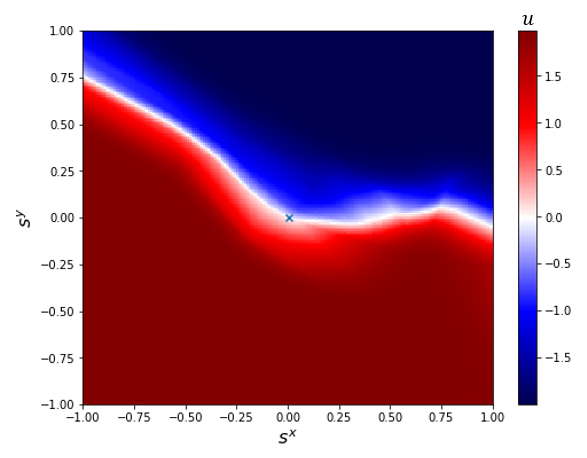}
	\caption{Learned control input at each state in the region $D'$. The color represents the value of the control input. The learned controller is not linear but nonlinear. The cross mark represents the convergence state $\bm{s}=[0.006,-0.001]^{T}$ ($\bm{x}=[1.000,0.000]^{T}$) in Fig.\ \ref{response}.\label{Controller} }
\end{figure}

Moreover, the time response of the controlled chaotic system by the controller that learned its control policy sufficiently is shown in Fig.\ \ref{response}, where the initial state is $\bm{x}_{0}=[0.2,1.8]^{T}$. It is shown that the controller inputs small control inputs when its state enters the region $D$ and stabilizes to the fixed point. Shown in Fig.\ \ref{Controller} is the control input at each state in the region $D$ by the learned controller. It is shown that the learned controller is not linear but nonlinear. Thus, the proposed method with continuous deep Q-learning can learn a nonlinear control policy for stabilizing a desired fixed point without identifying a mathematical model of the chaotic system.

\section{Conclusion and Future Work}
In this paper, we proposed the control method to stabilize a periodic orbit embedded in discrete-time chaotic system using DRL, where the model of the discrete-time system is not identified. Moreover, we show the usefulness of the proposed learning algorithm by the numerical simulation of the Gumowski-Mira map. It is future work to propose the chaos control method for continuous-time chaotic system with a P\'{o}incare map.


%
%

%

\begin{acknowledgments}
This work was partially supported by JST-ERATO HASUO Project Grant Number JPMJER1603, Japan and JST-Mirai Program Grant Number JPMJMI18B4, Japan.
\end{acknowledgments}


\end{document}